\documentclass[twocolumn,showpacs,prlprintnumbers,prl,fleqn,superscriptaddress]{revtex4}
\usepackage{amsfonts}
\usepackage{amssymb,amsbsy,amsmath,mathrsfs}
\usepackage[dvipdfmx]{graphicx}
\usepackage{bm}
\usepackage{subfigure}
\usepackage{color}
\makeatletter
\def\lsim{\mathrel{\mathpalette\gl@align<}}
\def\gsim{\mathrel{\mathpalette\gl@align>}}
\def\gl@align#1#2{\lower.6ex\vbox
{\baselineskip\z@skip\lineskip\z@
\ialign{$\m@th#1\hfil##\hfil$\crcr#2\crcr\sim\crcr}}}

\makeatother

\newcommand\ba{\begin{eqnarray}}
\newcommand\ea{\end{eqnarray}}
\newcommand\be{\begin{equation}}
\newcommand\ee{\end{equation}}
\newcommand\bi{\bibitem}
\newcommand{\ct}{\cite}

\def\non{\nonumber}

\def\de{\delta}

\def\ga{\gamma}

\def\la{\lambda}

\begin{document}

\title{Dynamics of decoherence: universal scaling of the decoherence factor}

\author{Sei Suzuki}
\affiliation{Department of Liberal Arts, Saitama Medical
University, Moroyama, Saitama 350-0495, Japan}

\author{Tanay Nag}

\author{Amit Dutta}
\affiliation{Department of Physics, Indian Institute of
Technology, Kanpur 208 016, India}
%
%
\begin{abstract}
We study the time dependence of the decoherence factor (DF) of a qubit  globally coupled   to an environmental spin system (ESS) which is driven across the
quantum critical point (QCP) by varying   a parameter of its Hamiltonian in time $t$ as $1 -t/\tau$ or $-t/\tau$, to which the qubit is coupled starting
at the time $t \to -\infty$; here, $\tau$ denotes the inverse quenching rate. In the limit of weak coupling, we analyze the time evolution of the DF in the vicinity of the
QCP (chosen to be at $t=0$) and define three quantities, namely, the generalized fidelity susceptibility $\chi_F(\tau)$ (defined right at the QCP), and the decay constants $\alpha_1 (\tau)$ and $\alpha_2 (\tau)$ which
dictate the decay of the DF at a  small but finite  $t$($>0$). Using a dimensional analysis argument based on the Kibble-Zurek
healing length, we show that   $\chi_F(\tau)$ as well as $\alpha_1 (\tau)$ and $\alpha_2(\tau)$ indeed satisfy universal
power-law scaling relations with  $\tau$ and the exponents  are solely determined by the spatial dimensionality
of the ESS and the exponents associated with its QCP. Remarkably, using the numerical t-DMRG method, these scaling relations are shown to be valid in both
the situations when the ESS is integrable and non-integrable and also for both linear and non-linear variation of the parameter. 
Furthermore, when an integrable ESS is quenched far away
from the QCP, there is a predominant Gaussian decay of the DF with a decay constant which also satisfies a universal scaling relation.


\end{abstract}
%
%
\maketitle



In the context of quantum computation and information \ct{nielson2000,vedral07}, one of the major issues  is the study of decoherence \ct{zurek03,joos03}, namely, the loss of coherence in a quantum system due to its interaction with the environment. To investigate the environment induced decoherence of a qubit in the vicinity of a quantum critical point (QCP)  \ct{sachdev99,suzuki13} of the environment, a paradigmatic  model known as the  central spin model (CSM) \ct{CSM1,CSM2} has been generalized
to the context of a quantum phase transition \ct{quan06}.

In the CSM a central spin (CS) or a qubit is globally coupled  to an environmental quantum many body system,
usually chosen to be a quantum spin system, referred to as the environmental spin system (ESS) in the subsequent discussions. The ESS is initially in its ground state while the CS is in a pure state; the global
coupling between the qubit and the environment is so chosen that the subsequent time evolution of the initial ground state
wave function of the ESS occurs along two channels dictated by two different Hamiltonians. Even though the qubit
is initially in a pure state, it has been shown that it loses its purity
(almost completely) \ct{quan06} when the ESS is close to its quantum
critical point (QCP). Question we address in this letter is as follows: what happens when the ESS
is slowly driven across its QCP? Is there a universality associated with the dynamically generated decoherence quantified
by the decoherence factor (DF) of the
CS especially in a limit when the coupling between the CS and the spin chain is weak?

\begin{figure}
\begin{center}
 \includegraphics[width=3.0in]{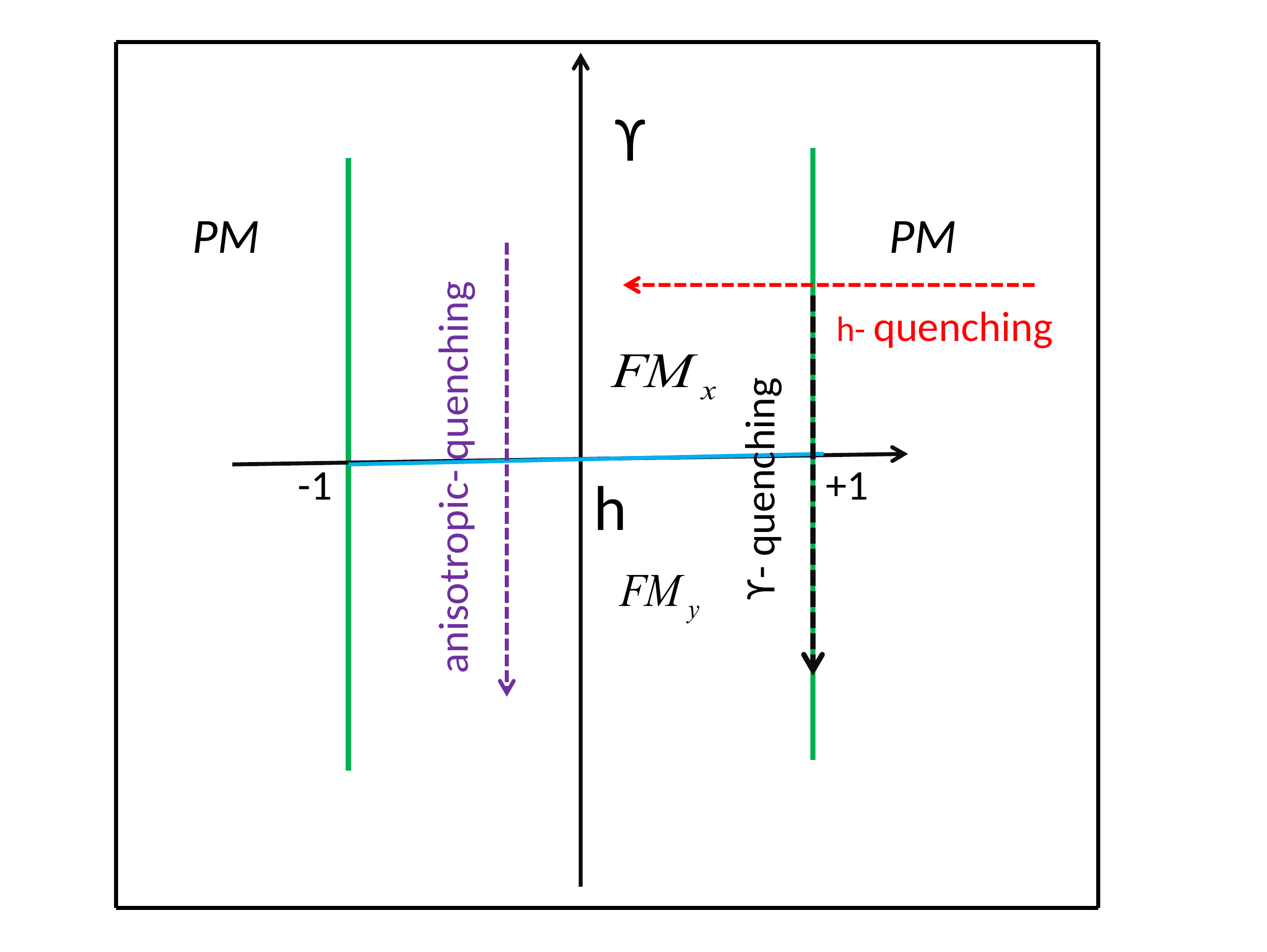}
 \end{center}
\caption{(Color online) The schematic phase diagram of the transverse XY chain given in Eq.~(\ref{eq_xy1}) in the $(h-\ga)$
 plane with different quenching paths denoted by arrows. The Ising transition line at $h=\pm 1$ for arbitrary $\ga$ and the anisotropic transition line at $\ga=0$ (with $|h| < 1$) meet at two mulcti-critical points (MCPs) $h=\pm 1$ and $\ga=0 $.  When the transverse field $h =1- t/\tau$ is quenched (e.g., setting $\gamma=1$) with $t$ from a large positive value to close to the QCP ($t >0$),
 one finds  $\chi_F (\tau) \sim \tau^{1/2}, \alpha_1(\tau) \sim \tau^0$, and $\alpha_2(\tau)\sim \tau^{-1/2}$ in the limit of 
 small $\delta$;
 identical scaling relations are obtained when
$\ga$ is quenched across the anisotropic critical line.
 Finally when $\ga$ is quenched linearly across the MCP with $h=1$,  one finds  $\chi_F(\tau)\sim\tau^{1/3}, \alpha_1(\tau)\sim\tau^{-1/3}$ and $\alpha_2\sim\tau^{-1}$. As shown in the text and in the supplementary material that all these scaling relations are predicted
 by  universal scaling form given in Eq.~ (\ref{eq:scaling_alpha}). 
  }
     \label{fig:xy_phase}
\end{figure}

Although we shall  consider more generic non-integrable  models in this letter,  let us first illustrate the basic idea using
the transverse XY chain consisting of $N$ spins  \ct{dutta15} as the environmental spin chain; the model is
described by the Hamiltonian 
\be
H_E^{\rm XY}=-\sum_{i=1}^N \left[J_x \sigma_i^x\sigma_{i+1}^x +J_y \sigma_i^y\sigma_{i+1}^y
+h \sigma_i^z\right] 
\label{eq_xy1}
\ee
where $\sigma$'s are the standard Pauli matrices, $J_x (J_y)$ is the ferromagnetic nearest neighbor interactions
along the $x$($y$) directions and $h$ is the transverse field;
this spin chain is coupled to the spin-1/2 qubit by a Hamiltonian $H_{SE}$. In the following, we shall set $J_x + J_y =1$
and the anisotropy in interaction i.e., $J_x-J_y$ will be denoted by the parameter $\gamma$.
The phase diagram and different phase transitions of the model (\ref{eq_xy1}) are presented in the Fig.~(\ref{fig:xy_phase}). 

Let us first assume that the qubit is coupled to the 
tunable transverse field of Eq.~(\ref{eq_xy1}) through the Hamiltonian
$H_{SE}=-\delta\sum_{i=1}^N\sigma_i^z\sigma_S^z$,
where $\sigma_i^z$ is the $i-$th spin of the XY chain and $\sigma_S^z$ represents that of the
qubit with $\delta$ being the coupling strength. [Notably, whenever the qubit is in the up (down) state, the field $h$ of the ESS gets altered to $h-\de$ ($h+\de$).] We choose
the qubit to be initially (at $t \to -\infty$)  in a pure state  
 $|\phi_S(t\to-\infty)\rangle= c_1|\uparrow\rangle+c_2|\downarrow\rangle$ with $|c_1|^2 + |c_2|^2=1$, where $|\uparrow\rangle$ and 
$|\downarrow \rangle$ represent up and down states of the CS, respectively, and the
environment is in the ground state $|\phi_E (t\to -\infty)\rangle = |\phi_g \rangle$.  The
initial  state of the composite Hamiltonian $H_E + H_{SE}$, at $t \to -\infty$, is then given by the direct product
$|\psi(t \to -\infty)\rangle=|\phi_S(t \to -\infty)\rangle \otimes |\phi_g\rangle$. 
It can be shown that at a later  time $t$, the composite wave function is given by
$|\psi(t)\rangle 
=c_1|\uparrow\rangle \otimes |\phi_+(t)\rangle + c_2|\downarrow \rangle \otimes |\phi_-(t)\rangle$,
where $|\phi_{\pm}\rangle$ are the wavefunctions evolving with the environment
Hamiltonian ${H^{\rm XY}_E(h\pm\delta)}$
given by the  Schr\"odinger equation 
$i {\partial}/{\partial t}|\phi_{\pm}\rangle =
{H^{\rm XY}_E(h\pm \delta)} |\phi_{\pm}\rangle.$
We therefore find that the coupling $\de$ essentially provides two channels of evolution of the environmental
wave function
dictated by two transverse XY Hamiltonians with the transverse field $h+\de$ and $h-\de$, respectively. It is straightforward to show that the DF 
defined through the relation $D(t)=|\langle
\phi_+(t)|\phi_-(t)\rangle|^2$ measures the purity of the reduced
density matrix of the qubit; any non-zero value of $\ln D(t)$ implies that the qubit is in a mixed state.
When the transverse field is close to the quantum critical value
there is a sharp dip in $D(t)$ which on the one hand, detects the existence of a QCP of the ESS and on the other, establishes that the qubit loses its initial purity almost completely in its vicinity \cite{quan06}.

The quantity $D(t)$ is also known as the Loschmidt echo which has been
studied in recent years (both at zero \ct{quan06} and finite temperatures \ct{zanardi07_echo})
in the context of decoherence in equilibrium \ct{quan06,rossini07,cucchietti07,venuti10,sharma12} and non-equilibrium situations \ct{damski11,nag12,mukherjee12,sharma14} and is closely connected to the dynamical phase transition \ct{heyl13,sharma15_dpt}, the statistics of work done
\ct{gambassi11} and  the  entropy generation in a quench \ct{dorner12,sharma15,russomanno15}. 

We assume a weak   coupling between
the qubit and  the environmental Hamiltonian; furthermore the qubit is always globally coupled to  time-dependent part of
the driven Hamiltonian. 
Under these circumstances, we now assume the generic situation
when a parameter of the ESS Hamiltonian is quenched $\la =-t/\tau$, with $t$ starting  from a large negative value; here,  $\la$ is the deviation from the  QCP (e.g., when the field $h$ in model (\ref{eq_xy1}) is quenched, $\la =h-1$), 
and $\tau (\gg 1) $ is the inverse quenching rate.
{Consequently, one has two channels of evolutions  of the initial ground state of the ESS dictated by two Hamiltonians with parameters $\la+\de$ and
$\la-\de$, respectively.}

\begin{figure}[t]
\begin{center}
 \includegraphics[width=8cm]{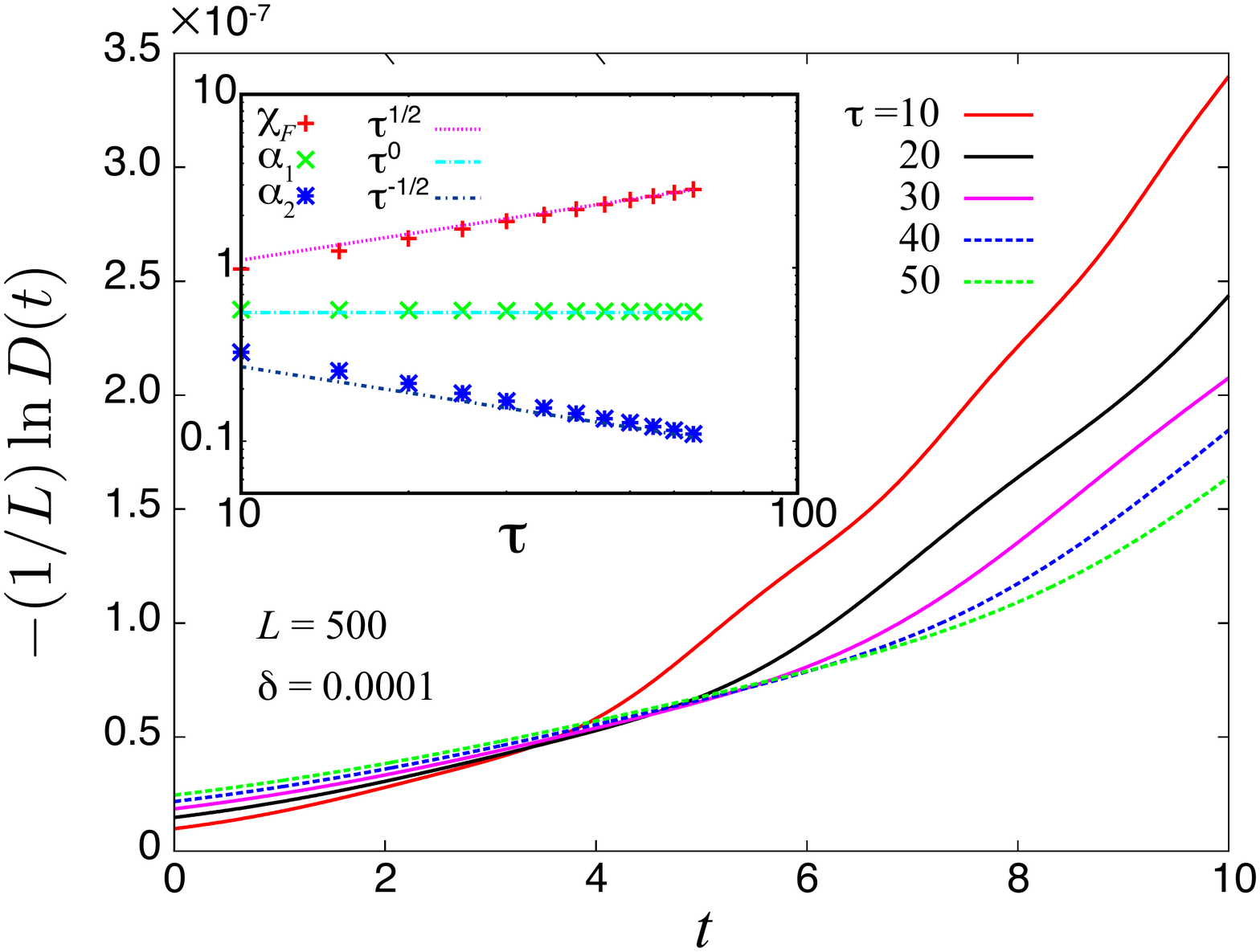}
\end{center}
\caption{Time evolution of the decoherence factor for the 
transverse Ising chain (TIC) as obtained using the t-DMRG method. The transverse field is 
driven as $h = 1 - t/\tau$. The scaling of $\chi_F(\tau), \alpha_1(\tau)$ and $\alpha_2(\tau)$ match nicely
with the theoretically predicted scaling relations given in Eq.~(\ref{eq:scaling_TIM}). }
\label{fig:TIM}
\end{figure}

In general, this quantity $D(t)$ can be expanded for sufficiently 
small $\delta$ and $t$ as
\begin{equation}
 D(t) \approx 1 - \left(\chi_F(\tau) + \alpha_1(\tau) t + 
\frac{1}{2}\alpha_2(\tau) t^2\right) \delta^2 L^d ,
\end{equation}
or equivalently,
\begin{equation}
 \frac{1}{L^d}\ln D(t) \approx
  - \left(
     \chi_F(\tau) + \alpha_1(\tau)t
     + \frac{1}{2}\alpha_2(\tau)t^2
    \right)\delta^2 ,
    \label{eq_decay}
\end{equation}
\noindent {with}
$$ \chi_F(\tau) = - \frac{1}{\delta^2 L^d}\ln D(0), ~~\alpha_m(\tau) = - \frac{1}{\delta^2 L^d} \frac{d^m}{dt^m}
  \left(\ln D(t)\right)|_{t=0},$$
where $L$ is the linear size of the system.
{Figure \ref{fig:TIM} shows the time evolution of $D(t)$ in the
{transverse Ising chain (TIC)} derived from  Eq.~(\ref{eq_xy1}) with $\gamma =1$ (i.e., $J_y=0$) driven as $h = 1 - t/\tau$. One can see that $D(t)$
follows Eq.~(\ref{eq_decay}) with $\chi_F$ and $\alpha_m$ changing with 
$\tau$.}
It should be noted that there is a prominent Gaussian early-time decay of $D(t)$ (i.e., $\alpha_1=0$)
as predicted in the Ref. [\onlinecite{peres84}] only when $D(t)$ is measured between the ground state
of unperturbed Hamiltonian; in the
present case  {$|\phi_+ (t=0)\rangle \neq |\phi_-(t=0)\rangle $}, 
and hence we indeed have a linear term in Eq.~(\ref{eq_decay}) which is also numerically established below and
analytically established  in the supplementary material (SM).

In this paper,  we present the universal scaling of these quantities namely,  the quantities $\chi_F(\tau)$, $\alpha_1 (\tau)$ and $\alpha_2(\tau)$.  We note that  $\ln D (\la=t=0) = {\ln ( |\langle \phi_{+}(t=0)|\phi_{-}(t=0)\rangle|^2)}$ in fact defines the fidelity \cite{comment2} between two states evolved
from the same initial state
along two different channels up to the value $\la=0$.
In the limit $\de \to 0$, $\ln D (\la=t=0)$ can be 
expanded in the form $\ln D (\la=t=0) = - \de^2 L^d \chi_F(\tau)$, where $\chi_F (\tau)$ can be viewed as a 
\textit {generalized fidelity susceptibility} (see the supplementary material (SM)).

In order to resolve the question raised above concerning the scaling of $\chi_F(\tau)$, $\alpha_1(\tau)$ and $\alpha_2(\tau)$, we resort to dimensional analysis.  When the ESS is slowly ramped across
its QCP, according to the Kibble-Zurek Scaling (KZS) \ct{zurek96,zurek2005,polkovnikov05} argument, the  non-adiabatic effects become prominent only in the vicinity
of the QCP; this is a consequence of the diverging
relaxation (healing) time at the QCP. As a result, there is a characteristic time
scale of the problem $\hat t \sim {\tau^{\nu z/(\nu z+1)}}$ where $\nu$ is the correlation length exponent and $z$ is
the dynamical exponent associated with the QCP across which the ESS is driven.  This follows from the simple adiabatic impulse argument which
relies on the assumption that at $t=\hat t$, the rate of change of the Hamiltonian equals the characteristic {(healing)} time of
the problem. So far as low energy modes are concerned, one can assume
that within the {time window} $-\hat t < t < \hat t$, known as the impulse region, the wave function remains frozen to the adiabatically evolved state at $t = -\hat t$, resulting in defects 
in the final state. One therefore finds a characteristic length scale, known as the healing length $\hat L$ that scales
as  $\hat L \sim \tau^{\nu /(\nu z +1)}$; assuming there is one defect within the healing length $\hat L$, one immediately finds the KZS
of the defect density $n$ in the final state following a slow ramp given by $n \sim 1/{\hat L}^d \sim \tau^{-\nu d/(\nu z +1)}$. We note that the proposed KZS form  has been verified and modified in several situations \ct{damski05,dziarmaga05,cherng06,Mukherjee2007,Sengupta2008,sen08,barankov08,Dutta2010,deng08,divakaran09} (For 
review see [\onlinecite{PolkovnikovRev,dziarmaga10,dutta15}]).
We shall demonstrate below how  the scaling of $\hat L$ leads us to the universal scaling of $\chi_F(\tau)$, $\alpha_1(\tau)$
and $\alpha_2(\tau)$
that we want to establish.

To derive the scaling, we first use the definition of the critical exponents 
$\nu$ and $z$ to find the scaling dimensions  $t \sim \hat L^z$ and $\de$ (which is equivalent to $\la$) $\sim \hat L^{-1/\nu}$.
Demanding that $\ln D (\la=t=0)$ must be dimensionless, we find 

\begin{eqnarray}
 \chi_F(\tau) &\sim& \tau^{(2 - d\nu)/(z\nu + 1)} ,\non\\~~
 \alpha_m(\tau) &\sim& {\tau^{(2 - d\nu - m z\nu)/(z\nu + 1)}}
  ~~~ (m = 1, 2) 
 \label{eq:scaling_alpha}.
\end{eqnarray}
Remarkably, the exponents appearing {in} scaling relations (\ref{eq:scaling_alpha}) are entirely determined by the universal critical
exponents and the spatial dimension $d$. As shown below, these relations
hold true for all integrable as well as  the non-integrable situations discussed here.  Remarkably, the scaling of $\chi_F(\tau)$ also
follows from the well established scaling of the fidelity susceptibility associated with the ground state fidelity (see the SM).

Question may arise what would happen if one considers a non-linear drive $\la (t) = {-|t/\tau|^{r}} {\rm sign}(t)$ of the ESS where ${\rm sign}(t)$ denotes the sign function. In
this case,   $\hat L$ scales as
$\tau^{r\nu /(r\nu z +1)}$ \ct{sen08,barankov08} and following a similar line of arguments, one finds

\begin{eqnarray}
 \chi_F(\tau) &\sim& \tau^{r(2 - d\nu)/(r z\nu + 1)} ,\non\\
 \alpha_m(\tau) &\sim& \tau^{r(2 - d\nu - m z\nu)/(r z\nu + 1)}
  ~~~ (m = 1, 2) .
  \label{eq:scaling_alpha_nonlin}
  \end{eqnarray}
Equations ({\ref{eq:scaling_alpha}) and (\ref{eq:scaling_alpha_nonlin}) contain the central result of our paper.

\begin{figure}[h!]
\centering
\subfigure[\  ]{
\includegraphics[width=4.1cm]{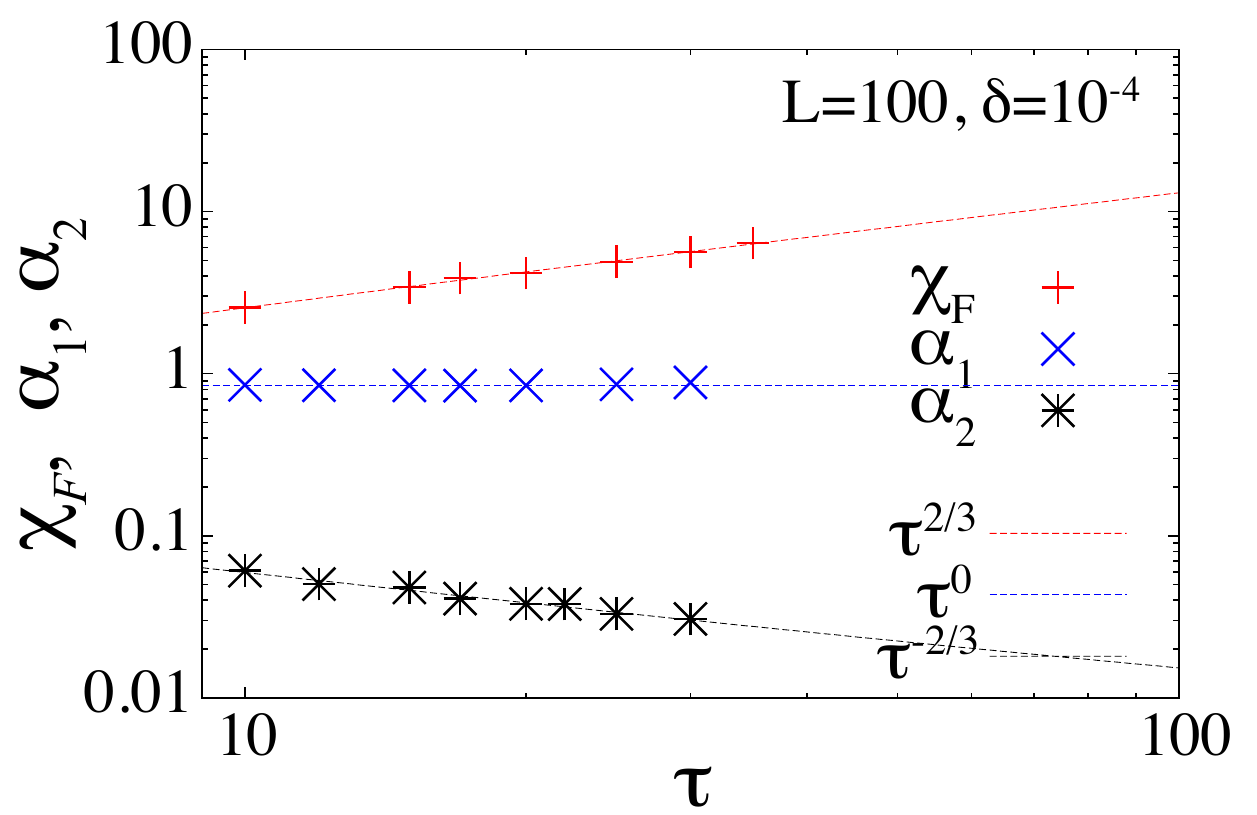}}
\subfigure[\ ]{
\includegraphics[width=4.1cm]{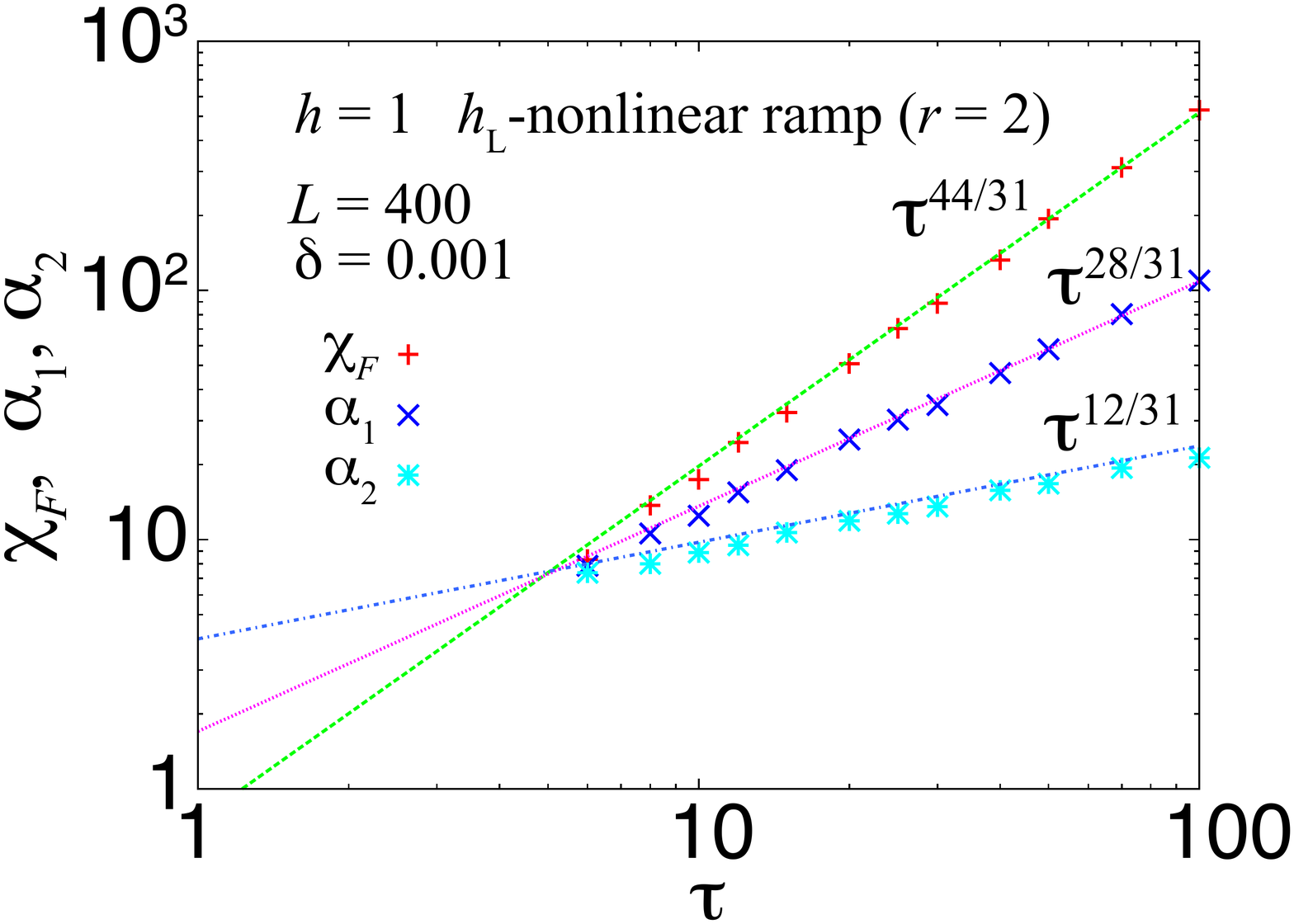}}
\caption{(Color online) (a) Scaling of $\chi_F$, $\alpha_1$, and $\alpha_2$
with respect to $\tau$ for the TIC (Eq.~(\ref{eq_xy1}) with $J_y=0$) with
a non-linear ramp ($r=2$) of $h$. We find an very good agreement with analytically predicted scaling in Eq.~(\ref{eq:scaling_TIM_nl}). 
(b) Scalings of $\chi_L$, $\alpha_1$ and $\alpha_2$
for the quadratic ramp ($r = 2$) of the 
longitudinal field in the non-integrable TIC given in (\ref{eq_ham_ferro})
with $h = 1$; the numerical results as obtained using t-DMRG are in excellent agreement with the predictions as in Eq.~(\ref{eq:NonInt-NL}).}
 \label{fig:non_lin}
 \end{figure}

%
%
\begin{figure}
 \includegraphics[width = 8cm]{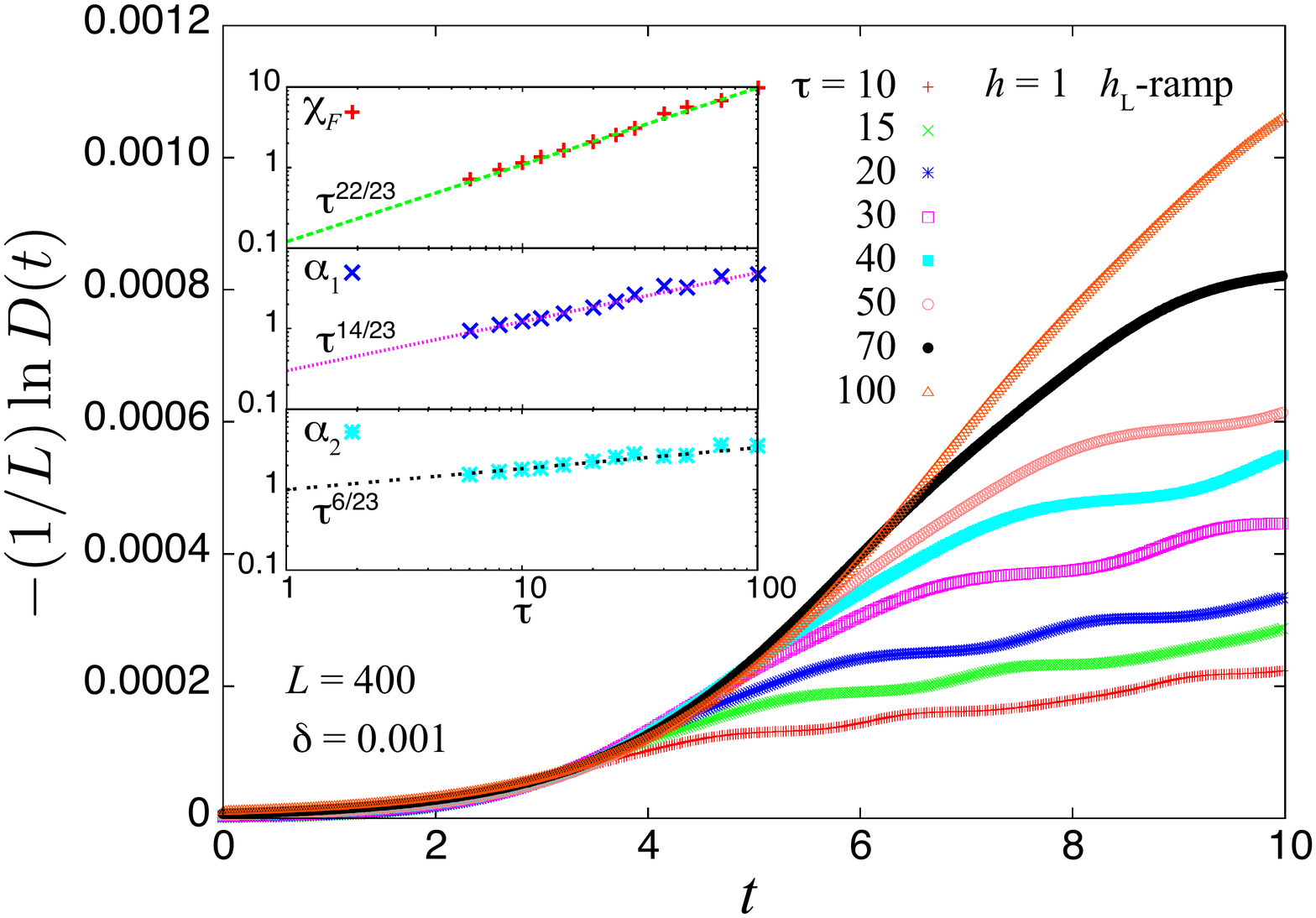}
\caption{
Early-time evolution of the decoherence factor
for the non-integrable TIC in the longitudinal
field (\ref{eq_ham_ferro}). The transverse field is fixed at $h = 1$ and the longitudinal
field is ramped as $h_L = -t/\tau$.
Inset shows scaling of $\chi_L$, $\alpha_1$ and $\alpha_2$.
Results are in good agreement with the prediction (\ref{eq:NonInt}).}
\label{fig:NonInt}
\end{figure}


Let us investigate above scaling relations for the integrable model (\ref{eq_xy1}) when the transverse field $h$ is quenched
as $h(t)= 1-t/\tau$ with $J_y=0$ and the spin chain is quenched across
(but close to) the QCP at $h=1$ (i.e., at $t=0$)
 with  exponent $z = \nu = d = 1$: Eqs.~(\ref{eq:scaling_alpha}) and (\ref{eq:scaling_alpha_nonlin}) predict: 
 \begin{equation}
 \chi_F(\tau)\sim \tau^{1/2},~~~ \alpha_1(\tau)\sim \tau^0, ~~~
\alpha_2(\tau) \sim \tau^{-1/2}
\label{eq:scaling_TIM}
\end{equation}
for the linear ramp, while
\begin{equation}
 \chi_F(\tau)\sim\tau^{r/(r+1)},~~~ 
\alpha_1(\tau) \sim\tau^{0}, ~~~ \alpha_2(\tau)\sim \tau^{-r/(r+1)}
\label{eq:scaling_TIM_nl}
\end{equation}
for the non-linear ramp. 
The scaling relations in Eq.~(\ref{eq:scaling_TIM}) has been verified numerically
(see Fig.~(\ref{fig:TIM})) and  established using exact analytical calculations
as shown in the SM. 
Scaling relations for the non-linear
ramp presented in {Equation (\ref{eq:scaling_TIM_nl})  is also numerically verified with an excellent agreement as shown
in Fig.~\ref{fig:non_lin}(a).
}

We now proceed to examine the universal scaling when ESS is non-integrable.  We  consider the situation when we have a ferromagnetic Ising chain in both transverse and longitudinal
fields \ct{pollmann10}:

\be
H_E^{\rm L} = -\sum_i \sigma_i^z \sigma_{i+1}^z - h_L\sum_i \sigma_i^z - \sum_i \sigma_i^x;
\label{eq_ham_ferro}
\ee
clearly the Hamiltonian $H_E^{\rm L}$ is at the quantum critical point when $h_L =0$ when it represents a critical quantum 
Ising Hamiltonian obtained by setting $h=J_x=1$ and $J_y=0$ in Eq.~(\ref{eq_xy1}). The dynamical exponent $z=1$ and a gap
opens up in the spectrum for small non-zero $h_L$ as $h_L^{8/15}$, leading to a value of $\nu=8/15$.
Considering a similar
coupling with the qubit of the form $H_{SE}=-\delta\sum_{i=1}^N\sigma_i^z\sigma_S^z$ with $\de \to 0$, we study the 
 quenching of the ESS close  the integrable quantum critical point ($h_L=0$) by varying $h_L=-t/\tau$ starting 
 from a large positive value of $h_L$.    Scaling relations  as obtained from Eqs.~(\ref{eq:scaling_alpha}) and
(\ref{eq:scaling_alpha_nonlin}) are
\begin{equation}
 \chi_F (\tau) \sim\tau^{22/23},
 \alpha_1(\tau)\sim\tau^{14/23},
 \alpha_2(\tau)\sim\tau^{6/23} ,
\label{eq:NonInt}
\end{equation}
for the linear ramp and
\begin{eqnarray}
 \chi_F(\tau)&\sim&\tau^{22r/(8r+15)},~~~
 \alpha_1(\tau)\sim\tau^{14r/(8r+15)},\non\\
 \alpha_2(\tau)&\sim&\tau^{6r/(8r+15)},
\label{eq:NonInt-NL}
\end{eqnarray}
for the non-linear ramp $h_L(t) = -|t/\tau|^r {\rm sign}(t)$.
Figure \ref{fig:NonInt} shows the early-time behavior of the
 DF
for the model (\ref{eq_ham_ferro}) following a linear quenching of
$h_L$ as obtained  using the t-DMRG method. 
The scalings of $\chi_F$, $\alpha_1$, and $\alpha_2$ extracted
close to $t=0$ are good
agreement  with the prediction in Eq.~(\ref{eq:NonInt}).
Figure \ref{fig:non_lin}(b), on the other hand,  shows the scaling of $\chi_F$,
$\alpha_1$, and $\alpha_2$ for the non-linear ramp; numerical results and theoretical prediction in 
Eq.~(\ref{eq:NonInt-NL}) with $r = 2$ are in excellent
agreement.

Let us now address the issue {that} what happens to the DF  when the system is quenched
far away from the QCP, i.e., $h$ in model (\ref{eq_xy1}) is changed from a large negative to a large positive value.
For an integrable model reducible to decoupled two-level problems (like the model (\ref{eq_xy1}) or the Kitaev honeycomb model \ct{kitaev06}), one can analytically establish that in the limit $t \gg 0$,
$\ln D \sim (-t^2 \de^2 L^d {\tilde\alpha}_2(\tau) )$, i.e., there is a prominent Gaussian decay with time \ct{damski11,nag12}. 
 Using the dimensional analysis argument presented above,
we immediately conclude that ${\tilde \alpha}_2(\tau) \sim \tau^{(2 - d\nu - 2 z\nu)/(z\nu + 1)}$ 
for a linear
quenching of $\lambda$ while for a non-linear ramp ${\tilde \alpha}_2(\tau) \sim  \tau^{r(2 - d\nu - 2 z\nu)/(r z\nu + 1)}$.
 It is noteworthy that whenever  $\nu z=1$, 
 the scaling of $\tilde{\alpha}_2$ is the same as that of $n$ for both $r=1$ and $r\neq1$; otherwise, the scaling
 of $n$ and ${\tilde \alpha}_2$ are completely different \ct{nag12}.
 The universal scaling formula of ${\tilde \alpha}_2 (\tau)$
explains all the integrable situations
discussed in earlier studies. (See the SM for relevant discussion.)

In summary, using the CSM and working in the weak coupling limit, we have provided a universal scaling relation of  the generalized fidelity susceptibility  $\chi_F(\tau)$ as well as the decay
constants $\alpha_1(\tau)$ and $\alpha_2(\tau)$ with the inverse rate $\tau$ in the vicinity of the QCP across which
the ESS is driven. We have  used a dimensional analysis arguments based on
the healing length $\hat L$. Remarkably,  the exponents are solely determined by the spatial dimension of the ESS and the associated
critical exponents. These scaling relations are verified for both integrable and non-integrable models as well as for
linear and non-linear ramps.
Additionally, we
have also derived a universal scaling relation for the decay constant ${\tilde \alpha_2}(\tau)$ describing the Gaussian decay of the DF with time when an integrable ESS
is quenched   far away from
the QCP.
{We reiterate that the DF is equivalent to the
Loschmidt echo and the latter has been experimentally observed in the
TIC using NMR technique
\ct{zhang09}. Therefore an experimental test of our scaling theory on
the DF can be realized in a large scale NMR quantum simulator. Furthermore, Kibble-Zurek scaling is already being explored in optical lattices \cite{braun15}. }

TN and AD acknowledge Uma Divakaran for fruitful discussions. AD  acknowledge SERB, DST India for financial support.

%

\vspace{-\baselineskip}

\widetext
\begin{center}
\textbf{\large Supplementary Material on ``Dynamics of decoherence: universal scaling of the decoherence factor''}\\
\vspace{0.5cm}
{ Sei Suzuki$^{1}$, Tanay Nag $^{2}$,  and Amit Dutta$^{2}$ }\\
\vspace{0.2cm}
{$^1$}{\it Department of Liberal Arts, Saitama Medical
University, Moroyama, Saitama 350-0495, Japan} \\
{$^2$}{\it Indian Institute of Technology Kanpur, Kanpur 208 016, India} \\
\end{center}

\setcounter{equation}{0}
\setcounter{figure}{0}
\setcounter{table}{0}
\setcounter{page}{1}
\makeatletter
\renewcommand{\theequation}{S\arabic{equation}}
\renewcommand{\thefigure}{S\arabic{figure}}
\renewcommand{\bibnumfmt}[1]{[S#1]}
\renewcommand{\@cite}[1]{[S#1]}

{\small In this supplementary material, we shall illustrate how to justify the expansion of $\ln D(t)$ close to $t=0$ and derive the scaling of $\chi_F (\tau)$, $\alpha_1(\tau)$ and $\alpha_2(\tau)$ for a transverse
Ising chain using scaling arguments and exact analytical calculations.

{
In order to estimate the DF in the limit $t \to 0+$, 
we first define the generalized time-dependent fidelity susceptibility
$\tilde{\chi}_F(t)$ as
\begin{equation}
 \tilde{\chi}_F (t) = - \frac{1}{\delta^2 L}\ln D (t) \big|_{\delta\to 0} ,
\end{equation}
and expend this quantity into Taylor series of $t$ as follows:
\begin{equation}
 \tilde{\chi}_F (t) = \chi_F(\tau) + \alpha_1(\tau)t + \frac{1}{2}\alpha_2(\tau)t^2 + \cdots ~.
\end{equation}
Therefore, the DF in the limit $t\to0+$ comes out to be
\begin{eqnarray}
D(t) &=& |\langle\phi_+(t)|\phi_-(t)\rangle|^2
 \simeq 1 -\delta^2 L \tilde{\chi}_F(t)  \\\nonumber
&\simeq& 1- \delta^2 L \biggl[\chi_F(\tau)+ \alpha_1(\tau)t + \frac{1}{2}\alpha_2(\tau)t^2\biggr].  
\label{eq:df}
\end{eqnarray}
In the following, we shall derive the scaling of $\chi_F(\tau)$,
$\alpha_1(\tau)$ and $\alpha_2(\tau)$ in integrable situations.
}}




\section{The scaling of $\chi_F (\tau)$, $\alpha_1(\tau)$ and $\alpha_2(\tau)$ for integrable situations }

We shall first derive the scaling of $\chi_F(\tau)$ or the \textit{generalized fidelity susceptibility} using the established scaling of the ground state quantum
fidelity.
As discussed in the main text, let us assume a linear quenching scheme  $\lambda=-t/\tau$ (for the quenching of
the transverse field $h$ in the transverse Ising
model, $\la =h-1$) which stops at {the} critical point at $ \la =t=0$.  We then calculate the logarithm  of the overlap or fidelity,  $\ln D (\la=t=0) = \ln ( |\langle \phi_{+}({t=0})|\phi_{-}({t=0})\rangle|)$;  expanding up to the second order in $\de$, one finds 
$\ln ( |\langle \phi_{+}({t=0})|\phi_{-}({t=0})\rangle|) \sim -\de^2 L^d \chi_F (\tau)$.  It is well established that at the QCP the ground
state quantum fidelity susceptibility scales as $L^{2/\nu -d}$ \cite{cite_fidelity}, where $\nu$ is the correlation length exponent, $d$ is the
spatial dimension of the system, and $L$ is the linear size of the system.  In the present problem of slow ramp, the characteristic length scale, as discussed in the 
main text, is given by the healing length $\hat L \sim \tau^{\nu/(\nu z +1)}$. Using $\hat L$ in the scaling of the ground state fidelity susceptibility one
arrives at the scaling relation $\chi_F(\tau) 
\sim \tau^{(2-\nu d)/(\nu z +1)}$, which has been obtained in the main text using a dimensional analysis approach. We note that
for this scaling to be valid, one must have  $\de^{-\nu} \gg L \gg \hat L$.

We shall illustrate  the above scaling considering the environmental  Hamiltonian to be a  transverse
XY chain
\begin{equation}
 H_{\rm E} = - \sum_{i=1}^L \left(
 J_x\sigma_i^x\sigma_{i + 1}^x + J_y \sigma_i^y \sigma_{i+1}^y + h(t)\sigma_i^z\right) ,
\label{eq:SM:H_E}
\end{equation}
and  we first consider the  Ising limit obtained by setting $J_y=0$; we also choose $J_x=1$.
The coupling Hamiltonian between the system
and the environment is given by
\begin{equation}
 H_{\rm SE} = - \delta \sum_{i=1}^L \sigma_i^z \sigma_S^z ,
\end{equation}
where $\sigma_i^z$ is the $i$-spin of the environment chain
and $\sigma_S^z$ represents the spin of the system.
Hereafter we assume a linear ramp of the transverse field:
\begin{equation}
 \la(t) = h(t) - 1 = - \frac{t}{\tau} ,
\end{equation}
where $\tau$ stands for the inverse rate of a ramp.

According to the exact solution presented in Ref.\cite{Sdamski11} for $L \to\infty$, the decoherence factor
is given  by
\begin{equation}
 D(t) \approx \exp\left(
  - \frac{L}{2\pi}\int_0^{\pi}
  dk \ln F_k^{-1}(t)
\right) ,
 \label{eq:D}
\end{equation}
\begin{equation}
 F_k(t) = |u_k^{+ \ast}(t) u_k^-(t) + v_k^{+ \ast}(t)v_k^-(t)|^2 ,
  \label{eq:F}
\end{equation}
where
$u_k^{\pm}$ and $v_k^{\pm}$ are the solutions of the 
time-dependent Bogoliubov-de Gennes equations
\begin{eqnarray}
 i\frac{d}{dt} u_k^{\pm}
  &=&  2 (h(t)\pm \delta - \cos k)u_k^{\pm} 
  + 2 \sin k \,v_k^{\pm} ,\\
 i\frac{d}{dt} v_k^{\pm}
  &=& 2 \sin k \,u_k^{\pm} 
  - 2 (h(t)\pm \delta - \cos k)v_k^{\pm} .
\end{eqnarray}
Using the notation,
$t_{\pm}' = - 4\tau \sin k (h(t)\pm \delta - \cos k)$,
$\tau' = 4\tau \sin^2 k$,
$z_{\pm} = t_{\pm}'e^{i\pi/4}/\sqrt{\tau'}$
and $n = -i\tau'/4$, the exact solution of the above
equations under the initial conditions $u_k^{\pm}(-\infty) = 1$
and $v_k^{\pm}(-\infty) = 0$ is obtained as
\begin{eqnarray}
 u_k^{\pm}(t) &=& e^{-\pi\tau'/16}e^{i\pi/4} \nonumber\\
 && \times \left[(1 + n)\mathcal{D}_{-n - 2}(iz_{\pm})
   + iz_{\pm}\mathcal{D}_{-n - 1}(iz_{\pm})\right], 
 \label{eq:u_ising}
\end{eqnarray}
\begin{equation}
 v_k^{\pm}(t) = \frac{\sqrt{\tau'}}{2}e^{-\pi\tau'/16}
  \mathcal{D}_{-n - 1}(iz_{\pm}) ,
  \label{eq:v_ising}
\end{equation}
where $\mathcal{D}_m(iz)$ is the Weber's parabolic 
cylinder function.
We note that $F_k(t) = 1$ when $\delta = 0$,
because $F_k(t)$ reduces to the norm of the wavefunction.

Hereafter we fix $t = 0$ so that $h(t) = 1$ and
assume $\tau \gg 1$.
Under this assumption, one can show that
only the long wave lengths with $k \lsim \tau^{-1/2} \ll 1$
give contribution to the integral of $\ln F^{-1}_k$
in Eq.~(\ref{eq:D}).
Therefore, in the present situation, one has
$t_{\pm}' \approx - 4\tau k (\pm \delta + k^2/2)$,
$\tau' \approx 4 \tau k^2$, 
$z_{\pm} \approx -2 \sqrt{\tau}(\pm\delta - k^2/2)$,
and $n \approx -i \tau k^2$.

Let us consider the situation where 
$\delta \ll \tau^{-1/2}\ll 1$.
Noting that $u_k^{\pm}$ and $v_k^{\pm}$
depend only on $\tau'$, $z_{\pm}$ and $n$,
one can see that
$\ln F_k^{-1}(0)$ is given as a function
of $\tau^{1/2}k$, $\tau^{1/2}k^2$ and $\tau^{1/2}\delta$.
Therefore the Taylor
expansion of $\ln F_k^{-1}(0)$ up to the order
of $\tau\delta^2$ leads to
$\ln F_k^{-1}(0) \approx \phi(\tau^{1/2}k, \tau^{1/2}k^2)
\tau\delta^2$,
where $\phi(x,y)$ is a function of $x$ and $y$.
We remark here that the linear term of $\delta$ vanishes
since $F_k$ is an even function of $\delta$.
To carry out the integral of $\ln F_k^{-1}$, 
one can safely shifts the upper limit to infinity because
there is no contribution to the integral
from $k\gsim \tau^{-1/2}$.
Consequently, one obtains
\begin{eqnarray}
 -\frac{1}{L}\ln D (\la=t=0) &\sim&
  \int_0^{\infty} dk\, \phi(\tau^{1/2}k,\tau^{1/2}k^2)\tau\delta^2
  \nonumber\\
 &\sim& \tau^{1/2}\delta^2\int_0^{\infty} d\tilde{k}\,
  \phi(\tilde{k},\tau^{-1/2}\tilde{k}^2) \nonumber\\
 &\sim& \tau^{1/2}\delta^2
\left\{\int_0^{\infty}\,d\tilde{k}\, \phi(\tilde{k},0) +
 \mathcal{O}\left(\tau^{-1/2}\right)
\right\} \nonumber\\
  &\sim& \tau^{1/2}\delta^2,
\end{eqnarray}
which gives $\chi_F (\tau)= -\ln D (\la=t=0)/(L \de^2) \sim \tau^{1/2}$.
Noting that the critical exponent $\nu$ associated with the QCP at $h=1$ is $\nu=1$ and spatial dimensionality $d=1$,
this scaling is in perfect congruence {with} the expected universal scaling behavior of $\chi_F$. In non-linear ramping case,
$\la = -|t/\tau|^r {\rm sign}(t)$, we can proceed using a similar arguments and establish that $\chi_F (\tau)  \sim \tau^{r/(r+1)}$.

Now, we shall calculate $F_k(t)$ explicitly by expanding the Weber
function, $\mathcal{D}_n(z)$, near $z = 0$.
One can show that $z\ll 1$ demands $t_{\pm}' \ll \sqrt{\tau'}$. We work in the limit of $\delta^2 \tau \ll 1$.
The power series expansion of $\mathcal{D}_n(z)$ for 
small $z$ is given by 
\begin{equation}
 \mathcal{D}_n(z)= \frac{2^{\frac{n}{2}} \sqrt{\pi}}{\Gamma(\frac{1}{2}-\frac{n}{2})}
 -\frac{2^{\frac{n+1}{2}} \sqrt{\pi} z}{\Gamma(\frac{-n}{2})}
 -\frac{2^{-2+\frac{n}{2}} \sqrt{\pi} (1+2n)z^2}{\Gamma(\frac{1}{2}-\frac{n}{2})}.
 \label{eq:weber}
\end{equation}
The solution of  time-dependent Bogoliubov-de Gennes equations for $t=\epsilon$ are given by
 \begin{eqnarray}
 &u_k^{\pm}(t)& \simeq e^{-\pi\tau'/16}e^{i\pi/4}\biggl[\frac{2^{-\frac{n}{2}} \sqrt{\pi}}{\Gamma(\frac{1}{2}+\frac{n}{2})}
 -\frac{\tau' z_{\pm} \sqrt{\pi}~ 2^{-\frac{n}{2}}}{4 \sqrt{2}~ \Gamma(1+\frac{n}{2})}
 +\frac{(1-2n)~z_{\pm}^2 \sqrt{\pi}~2^{-\frac{n}{2}}}{\Gamma(\frac{1}{2}+\frac{n}{2})}\biggr],
 \label{eq:upm}
 \end{eqnarray}
\begin{eqnarray}
 &v_k^{\pm}(t)& \simeq \frac{\sqrt{\tau'}}{2}e^{-\pi \tau'/16}~2^{-\frac{n}{2}}\sqrt{\frac{\pi}{2}} \biggl[\frac{1}{\Gamma(1+\frac{n}{2})} - \frac{\sqrt{2}~i z_{\pm}}{\Gamma(\frac{1}{2}+\frac{n}{2})}
 -\frac{(1+2n)~z_{\pm}^2}{4~\Gamma(1+\frac{n}{2})}\biggr] ,
 \label{eq:vpm}
 \end{eqnarray}
{where $z_{\pm}$ is defined by $z_{\pm} = (t_{\pm}'/\sqrt{\tau'})e^{i\pi/4}$.}
Using the above expressions one can find out the quantity $u_k^{+ \ast}(t) u_k^-(t) + v_k^{+ \ast}(t)v_k^-(t)$ which is given 
by 
\begin{eqnarray}
u_k^{+ \ast}(t) u_k^-(t) + v_k^{+ \ast}(t)v_k^-(t) &\simeq& 1+e^{-\pi \tau'/8}\biggl[z_+^*~a_1+z_-~a_2 +z_+^*z_-~a_3+(z_+^*)^2~a_4+z_-^2~a_5+z_+^*z_-^2~a_6+(z_+^*)^2 z_-~a_7 \biggr]  \nonumber \\
&& + O(z_{\pm}^4),
\label{eq:amp_F}
\end{eqnarray}
where 
\begin{eqnarray}
&a_1&= {i \pi\tau'\sqrt{2} \over 8~\Gamma({1\over 2}+{n^* \over 2})\Gamma(1+{n \over 2})} 
-{\pi \tau' \sqrt{2} \over 8~\Gamma({1\over 2}+{n \over 2})\Gamma(1+{n^* \over 2})}, \nonumber \\
&a_2&= a_1^*,\nonumber\\
&a_3&= {\pi \tau' \over 4~|\Gamma({1 \over 2}+{n \over 2})|^2} + {\pi \tau'^2 \over 32~ |\Gamma(1 +{n \over 2})|^2}, \nonumber \\
&a_4&=-{\pi \tau'(1+2n^*) \over 32 ~|\Gamma(1 +{n \over 2})|^2} + {\pi (1 -2n^*) \over |\Gamma({1 \over 2}+{n \over 2})|^2}, \nonumber \\
&a_5&=a_4^*,\nonumber \\
&a_6&=-{i \pi \tau'(1+2n)\sqrt{2} \over 32~ \Gamma({1\over 2}+{n^* \over 2})\Gamma(1+{n \over 2})}
-{\pi \tau'(1-2n) \over 4\sqrt{2}~\Gamma({1\over 2}+{n \over 2})\Gamma(1+{n^* \over 2}) },\nonumber \\
&a_7&=a_6^*.\nonumber \\
\label{eq:terms1}
\end{eqnarray}
Therefore, $F_k(t)$ is given by 
$F_k(t)=1+A_k(t)+B_k(t)+C_k(t)+D_k(t)$, where 
\begin{eqnarray}
A_k(t)&=&2 e^{-\pi\tau'/8}\biggl[|{\rm Re}(z_+a_1^*)|+|{\rm Re}(z_-a_2)|\biggr],\nonumber \\
B_k(t)&=&2 e^{-\pi\tau'/8}\biggl[|{\rm Re}(z_+^*z_-a_3)|+|{\rm Re}(z_+^2a_4^*)| +|{\rm Re}(z_-^2a_5)|\biggr] +e^{-\pi\tau'/4}\biggl[2|{\rm Re}(z_-z_+a_2a_1^*)|+|z_+|^2|a_1|^2+|z_-|^2|a_2|^2 \biggr],\nonumber \\
C_k(t)&=&2e^{-\pi\tau'/8}\biggl[ |{\rm Re}(z_+^*z_-^2a_6)|+|{\rm Re}(z_+^2z_-^*a_7^*)| \biggr]+2e^{-\pi\tau'/4}\biggl[ |{\rm Re}(|z_+|^2z_-a_1^*a_3)|+|{\rm Re}(|z_-|^2z_+a_2a_3^*)| \nonumber \\
&+&|{\rm Re}(z_+^*z_+^2a_1a_4*)| +|{\rm Re}(z_+^*(z_-^*)^2a_1a_5^*)|+ |{\rm Re}(z_-z_+^2a_2a_4^*)| +|{\rm Re}(z_-(z_-*)^2a_2a_5^*)| \biggr],\nonumber \\
D_k(t)&=&e^{-\pi \tau'/4}\biggl[ |z_+|^4|a_4|^2+|z_-|^4|a_5|^2 +|z_+|^2|z_-|^2|a_3|^2\biggr]+2e^{-\pi \tau'/4}\biggl[|{\rm Re}((z_+^*)^2(z_-^*)^2a_4a_5^*)|\nonumber \\
&+& |{\rm Re}(|z_+|^2(z_-^*)^2a_1a_6^*)|+ |{\rm Re}(z_+z_-^*|z_+|^2a_1a_7^*)|+ |{\rm Re}(z_+z_-^*|z_-|^2a_2a_6^*)| + |{\rm Re}(z_+^*z_-^*|z_-|^2a_3a_5^*)| \nonumber \\
&+& |{\rm Re}(z_+^2|z_-|^2a_2a_7^*)| + |{\rm Re}(z_+z_-|z_+|^2a_3a_4^*)| \biggr].\nonumber 
\label{eq:terms2}
\end{eqnarray}
The necessary algebra leads us to the conclusion that $A_k(t)=0$, which establishes the fact that there is no linear
 $\delta$ term present in $F_k(t)$. One can get the follwing contribution from $B_k(t)$
\begin{eqnarray}
 &B_k(t)&\simeq -\delta^2\biggl[ 2 \tau \tau' e^{-\pi \tau'/2} +
 3 \tau \tau' e^{-\pi \tau'/4}\biggr] -t^2 \biggl[ {2  \tau'  \over \tau} e^{-\pi \tau'/2} + {3  \tau'  \over \tau} e^{-\pi \tau'/4}\biggr] 
 \label{eq:term_b}
\end{eqnarray}
 Similarly, $C_k(t)$ gives $O(z_{\pm}^3)$ term,
\begin{eqnarray}
 &C_k(t)&\simeq-\delta^2  ~t~ \tau^{1/2} e^{-\pi \tau'/8}\biggl[{96 \pi \tau'^2\over 
64~ |\Gamma({1\over 2}+{n^* \over 2})\Gamma(1+{n \over 2})|} 
 - {96 \pi \tau'\over 8~|\Gamma({1\over 2}+{n \over 2})\Gamma(1+{n^* \over 2})|}\biggr] 
- {40~\delta^2 \pi t~ \tau^{1/2} \tau' e^{-3 \pi \tau'/8} \over 3
|\Gamma({1\over 2}+{n^* \over 2})\Gamma(1+{n \over 2})| }\nonumber \\
 &+& t^3  ~ \tau^{-3/2} e^{-\pi \tau'/8}
  \biggl[{96 \pi \tau'^2\over 
64~ |\Gamma({1\over 2}+{n^* \over 2})\Gamma(1+{n \over 2})|}
- {6 \pi \tau'\over |\Gamma({1\over 2}+{n \over 2})\Gamma(1+{n^* \over 2})|} \biggr]
- {5 t^3\tau^{-3/2}\pi \tau'  e^{-3 \pi \tau'/8} \over
|\Gamma({1\over 2}+{n \over 2})\Gamma(1+{n^* \over 2})|} \nonumber \\
\label{eq:term_c}
\end{eqnarray}
The fourth order term in $z_{\pm}$ is given by $D_k(t)$,
 \begin{eqnarray}
&D_k(t)&\simeq -\delta^2 t^2 \biggl[ e^{-\pi \tau'/2} \biggl( \alpha_1 +\beta_1 \tau'^2  \biggr) 
  +e^{-\pi \tau'/4} \times \biggl( \alpha_2 +\beta_2 \tau'^2  \biggr) \biggr]
-t^4  \biggl[ {e^{-\pi \tau'/2} \over \tau^2} \biggl( \alpha_3 + 
\eta_1 \tau' +\beta_3 \tau'^2\biggr)  \nonumber \\
 &+&  {e^{-\pi \tau'/4} \over \tau^2} \biggl( \alpha_4 + \eta_2\tau' +\beta_4 \tau'^2\biggr)\biggr] - \delta^4 \biggl[ e^{-\pi \tau'/2} \tau^2 \biggl( \alpha_5 + 
\eta_3 \tau' +\beta_5 \tau'^2\biggr) 
 +  e^{-\pi \tau'/4} \tau^2 \biggl( \alpha_6 + \eta_4\tau' +\beta_6 \tau'^2\biggr)\biggr],
 \label{eq:term_d}
\end{eqnarray}
where $\alpha$'s, $\beta$'s and $\eta$'s are the numbers coming from the $\Gamma$-function.

Using the approximation of  $\ln(1+x)\simeq x$ for $x\ll 1$ and expanding near the critical point $h=1$ and the critical mode
$k=0$, we can now compute the DF by evaluating the following integral over momentum space:
\begin{eqnarray}
&\int_0^{\infty}& dk \ln F_k(t) \simeq -\delta^2~(\alpha \tau^{1/2} + \beta t +\eta t^2 \tau^{-1/2}) - \phi t^2 \tau^{-3/2} -\rho t^3 \tau^{-2} -\gamma t^4 \tau^{-5/2} - \theta \delta^4 \tau^{3/2} \nonumber \\
&\simeq& -\delta^2~(\alpha \tau^{1/2} + \beta t +\eta t^2 \tau^{-1/2}) +\cdots
\end{eqnarray}
In the last line we have included only the leading order contributions to the integral. Here, $\alpha$, $\beta$, $\cdots$ are the numerical prefactors 
coming from the integartion of the dimensionless $\Gamma$-functions and the dimensionless exponentials.

The DF in the limit {$t=\epsilon$, $\epsilon \to +0$}, is therefore given by
\begin{eqnarray}
 D(t)&\approx& \exp\left( \frac{L}{2\pi}\int_0^{\pi} dk \ln F_k(t)\right)\nonumber \\
&\simeq& \exp\biggl(-{\delta^2 L \over 2\pi }~\biggl[\alpha \tau^{1/2} + \beta t +\eta t^2 \tau^{-1/2}\biggr]\biggr)\nonumber \\
&\simeq& 1- {\delta^2 L \over 2\pi }~\biggl(\alpha \tau^{1/2} + \beta t +\eta t^2 \tau^{-1/2}\biggr)
\end{eqnarray}
Clearly, one finds that $\chi_F(\tau) \sim \tau^{1/2}$, $\alpha_1(\tau)\sim \tau^0$ and $\alpha_2(\tau) \sim \tau^{-1/2}$ as
predicted by dimensional analysis argument given in Eq.~(\ref{eq:scaling_TIM}).

Referring to the Fig.~1 of the main text, we consider now the case when 
{$\gamma = J_x - J_y$ is ramped
as $\gamma=-t/\tau$ with $J_x + J_y = h = 1$ fixed in
Eq.~(\ref{eq:SM:H_E}). In this case, one has
a ramp along the gapless Ising critical line across the MCP.}
Using the notion of the dominant critical point  and
exponents $d = \nu = 1$ and $z = 2$\cite{Sdeng08,Sdivakaran09}, 
Eq.~(\ref{eq:scaling_alpha}) yields
\begin{equation}
  \chi_F(\tau)\sim\tau^{1/3}, ~~~ \alpha_1(\tau)\sim\tau^{-1/3}, ~~~
\alpha_2(\tau) \sim\tau^{-1}
\label{eq:SM:scaling_gapless}
\end{equation}
for the linear ramp (see Fig.~\ref{fig:SM:NonInt}(a) ) 
while
\begin{equation}
 \chi_F (\tau)\sim\tau^{r/(2r + 1)}, ~~~ \alpha_1 (\tau) \sim\tau^{-r/(2r+1)}, ~~~
\alpha_2(\tau)\sim\tau^{-3r/(2r+1)}
\label{eq_scaling_gapless_non_lin}
\end{equation}
for the non-linear ramp (see Fig.~\ref{fig:SM:NonInt}(b))  .


{We shall verify the above scaling form
of $\chi_F(\tau)$, $\alpha_1(\tau)$ and $\alpha_2(\tau)$
by explicitly calculating the DF.} The interaction Hamiltonian
between the environment and the qubit is accordingly modified as 
$H_{SE}=-(\delta/2)\sum_i(\sigma_i^x\sigma_{i+1}^x-\sigma_i^y\sigma_{i+1}^y)\sigma_S^z$. The
coupling $\delta$ therefore provides two channels of the
temporal evolution of the environmental ground state with anisotropy 
$\gamma +\delta $ and $\gamma-\delta$. 
The time-dependent Bogoliubov-de Gennes equation governing the time
evolution of environmental state
through the gapless line are  given by
\begin{eqnarray} 
 i\frac{d}{dt} u_k^{\pm}  &=&  2 (\gamma(t)\mp \delta ) ~
  \sin k ~u_k^{\pm}   + 2(h - \cos k) \,v_k^{\pm} ,\\
 i\frac{d}{dt} v_k^{\pm}  &=& 2 (h - \cos k) \,u_k^{\pm} 
  - 2 (\gamma(t)\mp \delta ) \sin k ~ v_k^{\pm} .
\end{eqnarray}
Accordingly, we use the notation,
$t_{\pm}' = 4\tau (h-\cos k)( \gamma(t)\mp \delta)$,
$\tau' = 4\tau (h-\cos k)^2 / \sin k$,
$z_{\pm} = t_{\pm}'e^{i\pi/4}/\sqrt{\tau'}$
and $n = -i\tau'/4$.

The solutions are the same as obtained for the case of transverse field
quenching given in 
Eq.~(\ref{eq:u_ising}) and Eq.~(\ref{eq:v_ising}).
In order to estimate $F_k(t)$ with {$t=\epsilon$, $\epsilon \to +0$}, one
has to use the following approximation: $t_{\pm}'\ll\sqrt{\tau'}$
and $\delta^2 \tau k \ll 1$. Following the same line of algebra one can
find out a similar expression of $u_k^{+ \ast}(t) u_k^-(t) + v_k^{+
\ast}(t)v_k^-(t)$
with a modified $z_{\pm}^{\rm gapless}=2\sqrt{\tau \sin k} (t/\tau~\mp
\delta)\exp(i\pi/4)$. One can note that for the Ising model
$z_{\pm}^{\rm Ising}=2\sqrt{\tau} (t/\tau~\mp \delta)\exp(i\pi/4)$. As a
result, $z_{\pm}^{\rm gapless}=\sqrt{\sin k}~z_{\pm}^{\rm Ising}$.
In this case, we find $A_k(t)=0$. The $O(z_{\pm}^2)$ contribution is
given by 
$B_k^{\rm gapless} = \sin k ~B_k^{\rm Ising}$. Similarly, $C_k^{\rm
gapless} = \sin^{3/2} k ~C_k^{\rm Ising}$ and
$D_k^{\rm gapless} = \sin^2 k ~D_k^{\rm Ising}$.

One can therefore calculate the integral of $F_k(t)$ by using the
logarithmic approximation and expanding momentum near the
critical mode, $\sin k \simeq k$ and $\tau'\simeq k^3
\tau$
\begin{eqnarray}
 &\int_0^{\infty}& dk \ln F_k(t) \simeq -\delta^2~(\alpha
  \tau^{1/3} + \beta t \tau^{-1/3}+\eta t^2 \tau^{-1}) -\phi t^2 \tau^{-5/3} -\rho t^3 \tau^{-7/3} -\gamma t^4 \tau^{-3} -
 \theta \delta^4 \tau \nonumber \\
&\simeq& -\delta^2~(\alpha \tau^{1/3} + \beta t \tau^{-1/3}+\eta t^2
 \tau^{-1}) + \cdots.
\label{eq:fk-gp}
\end{eqnarray}
The DF $D(t)$ in the vicinity of  MCP at {$t=\epsilon$, $\epsilon \to +0$}
 is given by
\begin{equation}
D(t)\simeq 1-{\delta^2 L \over 2\pi}~(\alpha \tau^{1/3} + \beta t
 \tau^{-1/3}+\eta t^2 \tau^{-1}) .
\end{equation}
{This expression of $D(t)$ clearly verifies the scaling of
$\chi_F(\tau)$, $\alpha_1(\tau)$ and $\alpha_2(\tau)$ predicted above
in Eq.~(\ref{eq:SM:scaling_gapless}).}

\begin{figure}[h!]
\centering
\subfigure[\  ]{
\includegraphics[width=6cm]{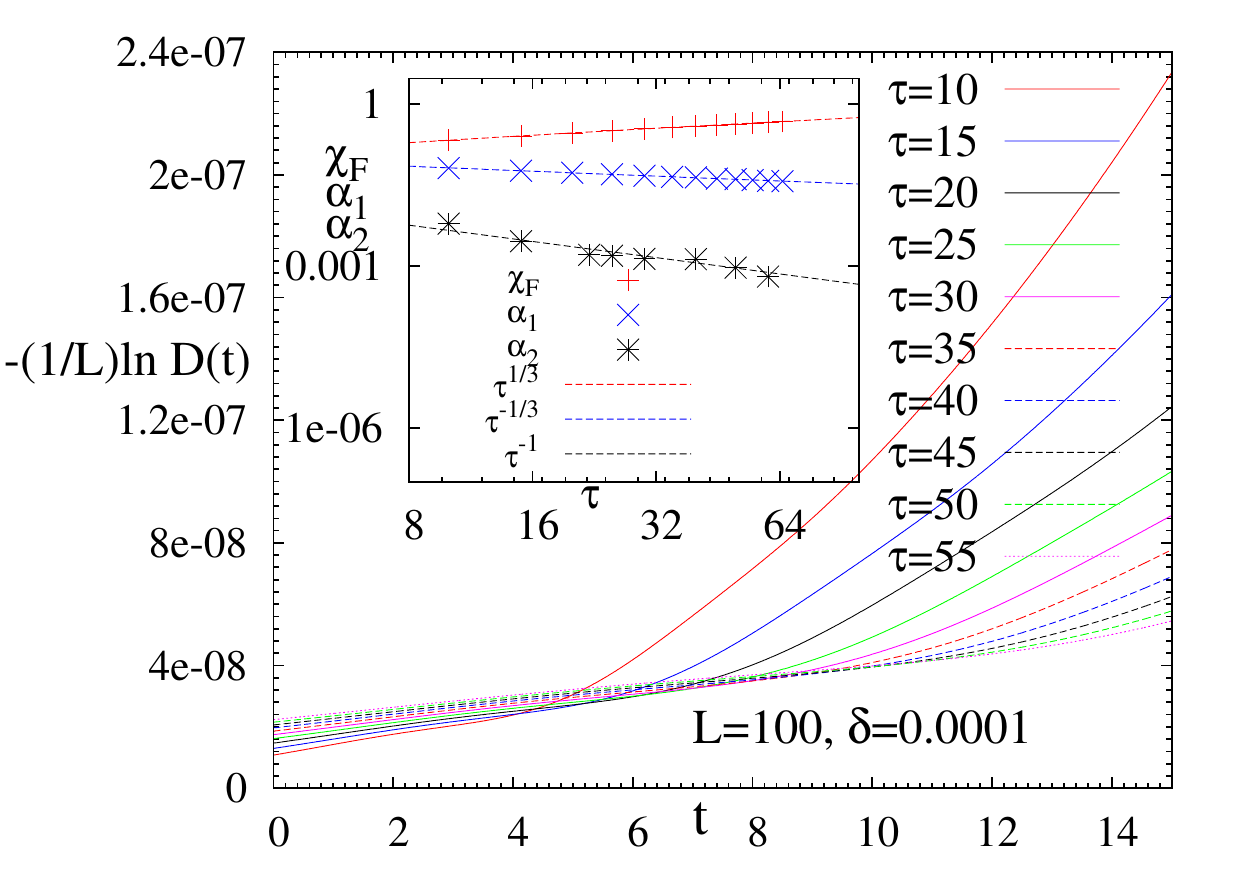}}
\subfigure[\ ]{
\includegraphics[width=6.5cm]{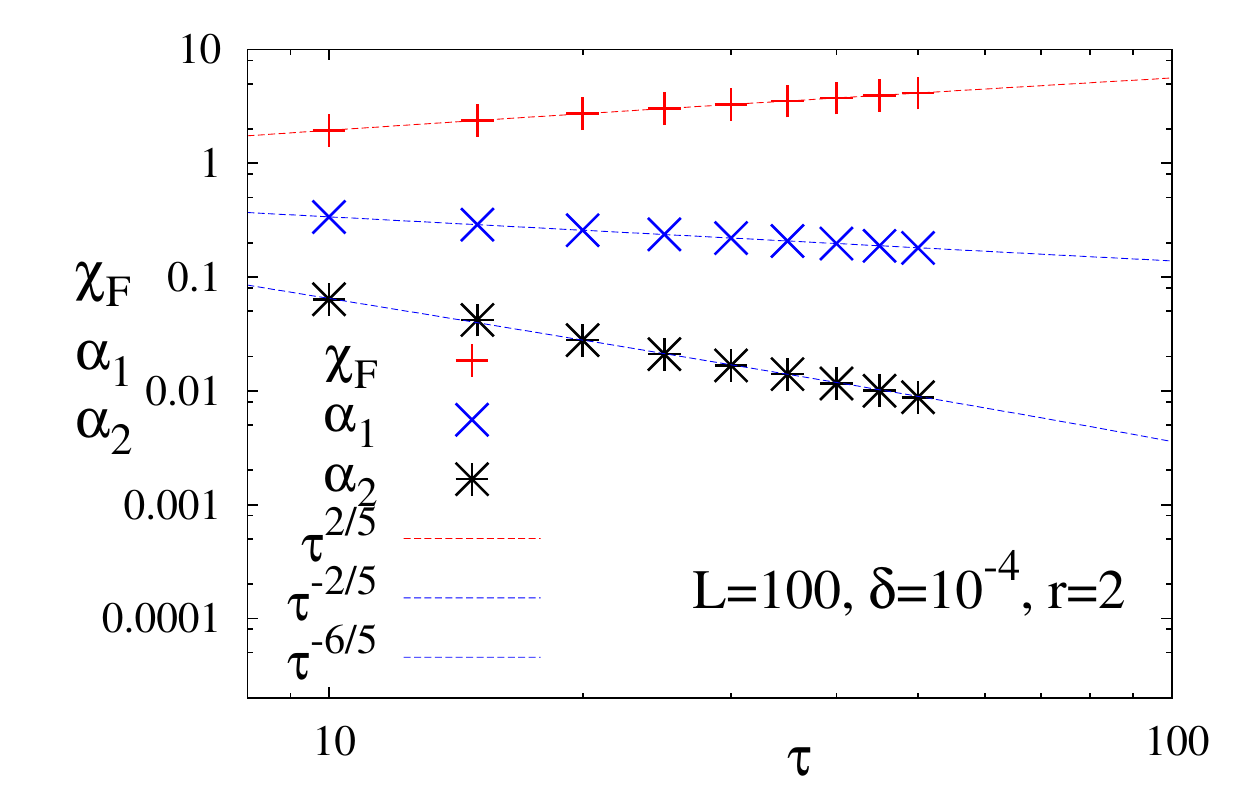}}
\caption{
(a) Early-time evolution of the decoherence factor
for quenching the parameter $\gamma$ as $-t/\tau$ of the Hamiltonian (\ref{eq_xy1})
with the transverse field is fixed at $h = 1$; the left panel  shows scaling of $\chi_L$, $\alpha_1$ and $\alpha_2$.
Results are in good agreement with the prediction in Eq.~(\ref{eq:SM:scaling_gapless}). (b) Numerical results  confirm
the scaling for the non-linear case given in Eq.~(\ref{eq_scaling_gapless_non_lin}).}
\label{fig:SM:NonInt}
\end{figure}

On the other hand, for an anisotropic quenching when $\gamma$ is quenched with $|h|<1$ and the gap vanishes for the critical
mode $k_c = \cos^{-1} h$, we have  $\tau' \simeq  4\tau \sqrt{1-h^2} k^2$ and $\sin k \simeq \sqrt{1-h^2}$; using these
parametrizations,  one
can retrieve scaling $\chi_F(\tau)$, $\alpha_1(\tau)$ and $\alpha_2(\tau)$ similar to those obtained
in the $h$-quenching case.

\section{Integrable models: scaling of $\ln D(t)$ far-away from the critical point $(t \gg 0$)}

Let us now address the issue {that} what happens to the DF  when the system is quenched
far away from the QCP, i.e., $h$ in model (\ref{eq_xy1}) is changed from a large negative to a large positive value.
For an integrable model reducible to decoupled two-level problems (like the model (\ref{eq_xy1}) or the Kitaev honeycomb model \ct{Skitaev06}), one can analytically establish that in the limit $t \gg 0$,
$\ln D \sim (-t^2 \de^2 L^d {\tilde\alpha}_2(\tau) )$, i.e., there is a prominent Gaussian decay with time \ct{Sdamski11,Snag12}. 
 Damski \textit{et al.} \ct{Sdamski11}, proposed that for an integrable ESS,  $\tilde{\alpha}_2 (\tau)$ is likely to 
scale as $n$ as determined by the KZS relation. In a subsequent work, Nag \textit{et al.} \ct{Snag12} showed that this conjecture is not necessarily true
even for the above integrable model when the system is quenched along a gapless critical line of the model (\ref{eq_xy1}); one
finds $n \sim \tau^{-1/3}$ while on the contrary ${\tilde \alpha}_2(\tau) \sim \tau^{-1}$.
 Using the dimensional analysis argument presented above,
we immediately conclude that ${\tilde \alpha}_2(\tau) \sim \tau^{(2 - d\nu - 2 z\nu)/(z\nu + 1)}$ 
for a linear
quenching of $\lambda$ while for a non-linear ramp ${\tilde \alpha}_2(\tau) \sim  \tau^{r(2 - d\nu - 2 z\nu)/(r z\nu + 1)}$.
 It is noteworthy that whenever  $\nu z=1$, 
 the scaling of $\tilde{\alpha}_2$ is the same as that of $n$. This holds true even for the non-linear ramp where both ${\tilde\alpha}_2$
 and $n$ scale as  $\tau^{-\frac{r \nu d }{(r \nu z +1)}}$. 
 On the other hand, when the environmental XY
chain is quenched  along the gapless Ising critical line $h=1$ (varying $\ga=t/\tau$ across the multi-critical point at $h=1,\ga=0$),
 one can argue that the multi-critical point plays the role of a dominant critical point with critical exponents
that $\nu=1$ and $z=2$ \ct{Sdeng08} and hence $\nu z \neq 1$; substituting these exponents in the KZS relation  and  the
predicted scaling ${\tilde \alpha}_2$, one immediately finds
 $n\sim \tau^{-1/3}$  and $\tilde \alpha \sim \tau^{-1}$.  Therefore, the universal scaling formula of ${\tilde \alpha}_2 (\tau)$
explains all the integrable situations
discussed in earlier studies.

\vspace{-\baselineskip}

\end{document}